\documentstyle[preprint,aps]{revtex}
\newcommand{\be}{\begin{equation}}
\newcommand{\ee}{\end{equation}}
\newcommand{\ba}{\begin{eqnarray}}
\newcommand{\ea}{\end{eqnarray}}

\begin{document}

\title{Black String Solutions with Arbitrary Tension}
\author{Chul H. Lee$^{}$ \footnote {e-mail: chulhoon@hanyang.ac.kr}}
\address{Department  of Physics, and BK21 Division of Advanced Research and Education in Physics, \\
Hanyang University, Seoul 133-791, Korea}
\maketitle

\begin{abstract}

We consider $1+4$ dimensional black string solutions which are
invariant under translation along the fifth direction. The
solutions are characterized by the two parameters, mass and
tension, of the source. The Gregory-Laflamme solution is shown to
be characterized by the tension whose magnitude is one half of the
mass per unit length of the source. The general black string
solution with arbitrary tension is presented and its properties
are discussed.

\end{abstract}

\newpage

\section{Introduction}

Spacetimes of dimensions higher than $1+3$ have become objects for
serious consideration in physics as some physical theories such as
the string theory and brane cosmology are necessarily formulated
in those higher dimensional spacetimes. In exploring the existence
of extra dimensions, the higher dimensional black hole solution
can be a useful tool. Along with the higher dimensional black
hole, another interesting object is the black string which is
obtained by extending the $1+3$ dimensional black hole to the
extra dimensions. The simplest case, first studied by Gregory and
Laflamme \cite{gre1}, is the black string obtained by extending
uniformly the Schwarzschild black hole to the fifth dimension. Its
metric is given by
\ba
 ds^{2}&=& g_{\mu \nu} dx^{\mu} dx^{\nu} \nonumber \\
 &=& -(1-\frac{r_{g}}{r}) dt^{2} + \frac{dr^{2}}{1-\frac{r_{g}}{r}}
 + r^{2} d\theta^{2} + r^{2} \sin^{2} \theta d\phi^{2} + dz^{2}.
 \label{schmetric}
\ea
Also Chamblin, Hawking and Real \cite{hawk} discussed black
strings in the models of Randall and Sundrum where our universe is
viewed as a domain wall in the $5$-dimensional anti-de Sitter
space.

The metric in Eq. (\ref{schmetric}) was shown in Ref. \cite{gre2}
to be unstable under small perturbations. The numerical analysis
there showed the existence of unstable modes for a range of time
frequency and wavelength. The possibiity of the unstable black
string finally fragmenting into black holes was mentioned.
However, Horowitz and Maeda \cite{horo} argued that event horizons
could not pinch off. So they conjectured that the unstable black
string solution would evolve to settle down to a new type of black
string solution which is not invariant under translation along the
string. This prompted many works on nonuniform black string
solutions (see \cite{kol} \cite{obe} for recent reviews). However
the discussions so far have been limited to perturbative or
numerical analyses; no analytic solutions have been found.

In this work we search for alternative uniform black string
solutions other than that of Eq. (\ref{schmetric}). We first note
that, as was discussed in \cite{ten1}, \cite{ten2}, two
independent asymptotic quantities are needed to characterize the
leading correction to the static metric far away from the source
on space $R^{d-1}\times S^{1}$. One of the quantities is the mass
and the other is the tension of the source. The metric in Eq.
(\ref{schmetric}) turns out to be, as is shown in the following,
the one produced by a source characterized by the tension of the
particular value of one half of the mass per unit length of the
source.

Let us restrict our discussion to cases of $1+4$ dimensional
spaces with the coordinates $x^{0}=t$, $x^{i}(i=1, 2, 3)$, and
$x^{4}=z$. For the weak gravitational field produced by stationary
$z$-independent weak source, the linearized Einstein equation in
harmonic coordinates is \be
 \partial_{i}\partial^{i} h_{\mu\nu} =  -16\pi G_{5}
 \overline{T}_{\mu\nu}  \label{lee}
\ee
where
\ba
  h_{\mu\nu} &=& g_{\mu\nu} - \eta_{\mu\nu}  \;\;\;
  (|h_{\mu\nu}|<< 1),  \\
 \overline{T}_{\mu\nu} &\equiv& T_{\mu\nu} - \frac{1}{3}\eta_{\mu\nu}T
\ea
and $G_{5}$ is the 5-dimensional gravitational constant. Using
the Green's function for the three dimensional Laplacian, the
solution of Eq, (\ref{lee}) can be written by
\be
 h_{\mu\nu}(x) = 4G_{5} \int d^{3}y \frac{\overline{T}_{\mu\nu}(y)}{|{\bf x} - {\bf y}|}
\ee
 With the expansion $\frac{1}{|{\bf x} - {\bf y}|} =
\frac{1}{r} + \frac{x^{i}y^{i}}{r^{3}} + \cdot\cdot\cdot$ where
$r=\sqrt{(x^{1})^{2}+(x^{2})^{2}+(x^{3})^{2}}$, the leading terms
of $h_{\mu\nu}(x)$ are then calculated to be
\ba
 h_{00}(x) &\simeq& \frac{4G_{5}}{r} \int d^{3}y (\frac{2}{3}T_{00}+\frac{1}{3}T_{44}) \nonumber \\
 h_{0i}(x) &\simeq& 2G_{5}\frac{x^{j}}{r^{3}}\int d^{3}y (y^{i}T^{0j}-y^{j}T^{0i}) \nonumber \\
 h_{04}(x) &\simeq& \frac{4G_{5}}{r}\int d^{3}y T_{04} \nonumber \\
 h_{ij}(x) &\simeq& \frac{4G_{5}}{3}\frac{\delta_{ij}}{r} \int d^{3}y (T_{00}- T_{44}) \nonumber \\
 h_{i4}(x) &\simeq& 2G_{5}\frac{x^{j}}{r^{3}}\int d^{3}y (y^{j}T^{4i}-y^{i}T^{4j}) \nonumber \\
 h_{44}(x) &\simeq& \frac{4G_{5}}{r}\int d^{3}y (\frac{1}{3}T_{00} + \frac{2}{3}T_{44})
\ea
 Here the relations $\int d^{3}y T^{i\mu} = 0$ and $\int d^{3}y y^{i}T^{\mu j} = -\int
d^{3}y y^{j} T^{\mu i}$, which are derived from the
energy-momentum conservation
 $\partial_{\nu}T^{\mu \nu} = \partial_{i}T^{\mu i} = 0$ (and $\partial_{k}(x^{i}T^{\mu k}) =
 T^{\mu i}$, $\partial_{k}(x^{i}x^{j}T^{\mu k}) = x^{j}T^{\mu i} + x^{i}T^{\mu j}$) are used.
For the static case with $T^{0i}=T^{04}=0$, the leading
corrections to the metric far away from the source, up to the
order of $\frac{1}{r}$, can be seen to be characterized by the two
quantities
\be
 \lambda \equiv \int d^{3}x T_{00}, \;\;  \tau \equiv -\int d^{3}x T_{44}.
\ee
 $\lambda$ is the mass per unit length and
$\tau$ is the tension along the $z$-direction of the source. Then,
up to the order of $\frac{1}{r}$,
\ba
 h_{00}(x) &\simeq& \frac{4G_{5}}{3} \frac{2\lambda - \tau}{r}
 \nonumber \\
 h_{ij}(x) &\simeq& \frac{4G_{5}}{3}\delta_{ij} \frac{\lambda + \tau}{r}
 \nonumber \\
 h_{44}(x) &\simeq& \frac{4G_{5}}{3} \frac{\lambda - 2\tau}{r}
 \label{hmn1}
\ea
 and all other components are zero.
If the direction $z$ is periodic with $0\leq z<L$, the four
dimensional gravitational constant $G_{4}$ is given by
$G_{4}=\frac{G_{5}}{L}$ and the total mass of the source is
$M=\lambda L$.

In section II, we show that the source of the Gregory-Laflamme
metric is characterized by the tension whose magnitude is equal to
one half of the mass per unit length. A general class of black
string solutions with arbitrary tension is presented in section
III. The spacetime properties of the new solutions are discussed
in section IV.

\section{The Characteristics of the Source of the Gregory-Laflamme Metric}

In order to examine the characteristics of the source that
produces the metric given in Eq. (\ref{schmetric}), one needs to
reexpress it in harmonic coordinates. We first replace $r$ with
$\rho$ defined by \be
 r\equiv \rho(1+\frac{r_{g}}{4\rho})^{2}
\ee
 Then the metric is rewritten by
\be
 ds^{2} = -\frac{(1-\frac{r_{g}}{4\rho})^{2}}{(1+\frac{r_{g}}{4\rho})^{2}} dt^{2}
  + (1+\frac{r_{g}}{4\rho})^{4} (d\rho^{2}+ \rho^{2} d\theta^{2}
  + \rho^{2} \sin^{2} \theta d\phi^{2}) + dz^{2}
  \label{gremetric2}
\ee
 And finally with the definition
\be
 x^{1} = \rho \sin \theta \cos \phi, \;\; x^{2} = \rho \sin \theta \sin \phi, \;\;
 x^{3} = \rho \cos \theta
\ee
 the metric becomes
\be
 ds^{2} = -\frac{(1-\frac{r_{g}}{4\rho})^{2}}{(1+\frac{r_{g}}{4\rho})^{2}} dt^{2}
  + (1+\frac{r_{g}}{4\rho})^{4} \delta_{ij}dx^{i} dx^{j} + dz^{2}
\ee
 A straightforward calculation shows that this form of the metric satisfies the
linearized harmonic coordinate condition,
$\partial_{\mu}h^{\mu}_{\;\; \nu} - \frac{1}{2} \partial_{\mu}h =
0$, up to the second order in $\frac{1}{\rho}$. The leading
corrections of the metric in the asymptotic region can now be read
to be
\be
 h_{00} = \frac{r_{g}}{\rho}, \;\; h_{ij} = \frac{r_{g}}{\rho}\delta_{ij}, \;\;
 h_{44} = 0 \label{hmn2}
\ee
Comparing Eq. (\ref{hmn2}) with Eq. (\ref{hmn1}) gives
\be
 \lambda = \frac{r_{g}}{2G_{5}}, \;\;\; \tau = \frac{r_{g}}{4G_{5}}
\ee
 Therefore one can conclude that the source of the Gregory-Laflamme metric is characterized by
the tension whose magnitude is one half of the mass per unit
length($\tau = \frac{1}{2}\lambda$).

\section{A General Class of Black String Solutions}

We now search for the metric whose source has some arbitrary
tension. Let us say the tension and the mass per unit length of
the source are related by
\be
 \tau = a\lambda
\ee
where $a$ is some arbitrary constant. Then we know, from Eq.
(\ref{hmn1}), that the leading corrections of the metric far away
from this source are
\be
 h_{00} = \frac{B(2-a)}{\rho} , \;\; h_{ij} = \delta_{ij}\frac{B(1+a)}{\rho},
 \;\;  h_{44} = \frac{B(1-2a)}{\rho}
\ee
 where $B=\frac{4G_{5}\lambda}{3}$. That is, the asymptotic form of
the metric is
\ba
 ds^{2} \approx &-& (1- \frac{B(2-a)}{\rho}) dt^{2}
 + (1+\frac{B(1+a)}{\rho}) (d\rho^{2}+ \rho^{2} d\theta^{2}
  + \rho^{2} \sin^{2} \theta d\phi^{2}) \nonumber \\
  &+& (1+\frac{B(1-2a)}{\rho} )dz^{2}
 \label{asymsol}
\ea
 We finally have to find the solutions of the vacuum Einstein
field equations which reduce to the asymptotic form of Eq.
(\ref{asymsol}) at large $\rho$. We start with the ansatz
\be
 ds^{2} = -F(\rho) dt^{2} + G(\rho)(d\rho^{2}+ \rho^{2} d\theta^{2}
  + \rho^{2} \sin^{2} \theta d\phi^{2}) + H(\rho) dz^{2},
  \label{genmet}
\ee
 and substitute this form of metric into the vacuum Einstein
field equations to derive differential equations for the three
functions $F(\rho)$, $G(\rho)$ and $H(\rho)$. The solutions turn
out to be
\ba
 F &=& (1-\frac{K_{a}}{\rho})^{s}(1+\frac{K_{a}}{\rho})^{-s} \nonumber \\
 G &=& (1-\frac{K_{a}}{\rho})^{2-\frac{1+a}{2-a}s}(1+\frac{K_{a}}{\rho})^{2+\frac{1+a}{2-a}s}
  \nonumber \\
 H &=& (1-\frac{K_{a}}{\rho})^{-\frac{1-2a}{2-a}s}(1+\frac{K_{a}}{\rho})^{\frac{1-2a}{2-a}s}
 \label{sol}
\ea
where
\be
 s = \frac{2(2-a)}{\sqrt{3(1-a + a^{2})}}, \;\;
 K_{a} =\sqrt{\frac{1-a + a^{2}}{3}} G_{5}\lambda
 \label{sk}
\ee
 For $a=\frac{1}{2}$, these solutions in Eq's (\ref{sol}) and (\ref{sk}) can be
seen to give Eq. (\ref{gremetric2}), the Gregory-Laflamme metric,
with $r_{g}=4K_{1/2}=2G_{5}\lambda=2G_{4}M$. For $a=0$(no
tension), the metric becomes
\ba
 ds^{2} = &-& \frac{(1-\frac{K_{0}}{\rho})^{4/\sqrt{3}}}
                   {(1+\frac{K_{0}}{\rho})^{4/\sqrt{3}}}dt^{2}+
              (1-\frac{K_{0}}{\rho})^{2-\frac{2}{\sqrt{3}}}
                   (1+\frac{K_{0}}{\rho})^{2+\frac{2}{\sqrt{3}}}
 (d\rho^{2}+ \rho^{2} d\theta^{2} + \rho^{2} \sin^{2} \theta d\phi^{2}) \nonumber \\
 &+& \frac{(1+\frac{K_{0}}{\rho})^{2/\sqrt{3}}}{(1-\frac{K_{0}}{\rho})^{2/\sqrt{3}}} dz^{2}
 \label{sol0}
\ea
  with $K_{0}=\frac{1}{\sqrt{3}}G_{5}\lambda$.
For $a=1$, the metric becomes
\ba
 ds^{2} = &-& \frac{(1-\frac{K_{1}}{\rho})^{2/\sqrt{3}}}
                 {(1+\frac{K_{1}}{\rho})^{2/\sqrt{3}}}dt^{2}+
            \frac{(1+\frac{K_{1}}{\rho})^{\frac{4}{\sqrt{3}}+2}}
                 {(1-\frac{K_{1}}{\rho})^{\frac{4}{\sqrt{3}}-2}}
 (d\rho^{2}+ \rho^{2} d\theta^{2} + \rho^{2} \sin^{2} \theta d\phi^{2}) \nonumber \\
 &+& \frac{(1-\frac{K_{1}}{\rho})^{2/\sqrt{3}}}{(1+\frac{K_{1}}{\rho})^{2/\sqrt{3}}} dz^{2}
 \label{sol1}
\ea
  with $K_{1}=\frac{1}{\sqrt{3}}G_{5}\lambda$.

\section{Spacetime Properties of the New Solutions}

Using, instead of $\rho$, a new coordinate $r$ defined by
\be
 r = \rho G^{1/2}(\rho)
\ee
 the metric in Eq. (\ref{genmet}) is transformed to
\be
 ds^{2} = -F dt^{2} + \frac{dr^{2}}{(1+\frac{\rho}{2G} \frac{dG}{d\rho})^{2}}
 + r^{2} (d\theta^{2} + \sin^{2} \theta d\phi^{2}) + H dz^{2}
\ee
 One can see that the event horizon is formed at the surface
satisfying the condition
\be
 1+\frac{\rho}{2G} \frac{dG}{d\rho} = 0,
\ee
or explitly
\be
 (\frac{\rho}{K_{a}}- \frac{1+a}{\sqrt{3(1-a+a^{2})}})^{2}
 +\frac{(1-2a)(2-a)}{3(1-a+a^{2})}=0
 \label{horcon}
\ee
For $a=1$, Eq. (\ref{horcon}) becomes
\be
  (\frac{\rho}{K_{1}}- \frac{2}{\sqrt{3}})^{2} - \frac{1}{3} = 0,
\ee
and the solution is
\be
 \rho=\sqrt{3}K_{1}=G_{5}\lambda
\ee
 or
\be
 r=\frac{(1+\frac{1}{\sqrt{3}})^{\frac{2}{\sqrt{3}}+1}}{(1-\frac{1}{\sqrt{3}})^{\frac{2}{\sqrt{3}}-1}}G_{5}\lambda
\ee
 For $a=\frac{1}{2}$,  Eq. (\ref{horcon}) becomes
\be
  (\frac{\rho}{K_{1/2}}- 1)^{2} = 0
\ee
 with $K_{1/2}=\frac{1}{2}G_{5}\lambda$, and the solution is
\be
 \rho=K_{1/2}=\frac{1}{2}G_{5}\lambda
\ee
 or
\be
 r=4K_{1/2}=2G_{5}\lambda
\ee
 For $a=0$,  Eq. (\ref{horcon}) becomes
\be
  (\frac{\rho}{K_{0}}- \frac{1}{\sqrt{3}})^{2}+\frac{2}{3} = 0,
\ee
and we see there is no solution to it. Therefore no event horizon
is formed in this case. In fact, for $a<\frac{1}{2}$
\be
 \frac{(1-2a)(2-a)}{3(1-a+a^{2})}>0,
\ee
and there is no solution to Eq. (\ref{horcon}). We conclude that
the event horizon is formed only when the tension is larger than
or equal to one half of the mass per unit length of the source.

We now concentrate on the case of no tension($a=0$) in the source
string. In this case, as was shown above, no event horizon is
formed. However there appears a spacetime singularity at $\rho =
K_{0}$($r=0$). This conclusion is confirmed by showing the
existence of singular components of the Riemann tensor with
respect an orthonormal frame. With the basis vectors of the
orthonormal frame taken to be
\ba
 e_{\hat{t}}^{\;\; \mu} &=&
 \frac{(1+\frac{K_{0}}{\rho})^{\frac{2}{\sqrt{3}}}}{(1-\frac{K_{0}}{\rho})^{\frac{2}{\sqrt{3}}}}\delta^{\mu}_{\;\; t},
  \;\; e_{\hat{\rho}}^{\;\; \mu} =
 \frac{\delta^{\mu}_{\;\; \rho}}{(1-\frac{K_{0}}{\rho})^{1-\frac{1}{\sqrt{3}}}(1+\frac{K_{0}}{\rho})^{1+\frac{1}{\sqrt{3}}}},
  \;\; e_{\hat{\theta}}^{\;\; \mu} =
 \frac{\delta^{\mu}_{\;\; \theta}}{\rho (1-\frac{K_{0}}{\rho})^{1-\frac{1}{\sqrt{3}}}
 (1+\frac{K_{0}}{\rho})^{1+\frac{1}{\sqrt{3}}}},
 \nonumber \\
  e_{\hat{\phi}}^{\;\; \mu} &=&
  \frac{\delta^{\mu}_{\;\; \phi}}{\rho \sin \theta (1-\frac{K_{0}}{\rho})^{1-\frac{1}{\sqrt{3}}}
 (1+\frac{K_{0}}{\rho})^{1+\frac{1}{\sqrt{3}}}}, \;\;
  e_{\hat{z}}^{\;\; \mu} =
 \frac{(1-\frac{K_{0}}{\rho})^{1/\sqrt{3}}}{(1+\frac{K_{0}}{\rho})^{1/\sqrt{3}}}\delta^{\mu}_{\;\;z},
\ea
 a straightforward calculation gives
\ba
 R_{\hat{t} \hat{\rho} \hat{t} \hat{\rho}}&=& \frac{8K_{0}(1-\frac{\sqrt{3}K_{0}}{\rho}+\frac{K_{0}^{2}}{\rho^{2}})}
 {\sqrt{3}\rho^{3}(1-\frac{K_{0}}{\rho})^{4-\frac{2}{\sqrt{3}}}(1+\frac{K_{0}}{\rho})^{4+\frac{2}{\sqrt{3}}}} \nonumber \\
 R_{\hat{t} \hat{z} \hat{t} \hat{z}}&=& - \frac{2K_{0}^{2}}
 {3\rho^{4}(1-\frac{K_{0}}{\rho})^{4-\frac{2}{\sqrt{3}}}(1+\frac{K_{0}}{\rho})^{4+\frac{2}{\sqrt{3}}}} \nonumber \\
 R_{\hat{\rho} \hat{z} \hat{\rho}\hat{z}}&=& \frac{4K_{0}(1+\frac{K_{0}^{2}}{\rho^{2}})}
 {\sqrt{3}\rho^{3}(1-\frac{K_{0}}{\rho})^{4+\frac{2}{\sqrt{3}}}(1+\frac{K_{0}}{\rho})^{4-\frac{2}{\sqrt{3}}}}.
\ea
 They are indeed singular at $\rho = K_{0}$($r=0$). Here we
seen an explicit example in which the cosmic censorship hypothesis
is not valid in the higher dimensional spacetime.

 The metric of Eq. (\ref{sol0}) is the one generated by a string
source with no tension located at $\rho = K_{0}$($r=0$). The
relevant range of the coordinate $\rho$ ($r$) is $K_{0} \leq \rho
<\infty$ ($0\leq r<\infty$). When the source string has a finite
thickness, the location  $\rho = K_{0}$($r=0$) is within the
source and the spacetime singularity does not materialize.

We leave the similar discussions for the general case of no event
horizon($a<\frac{1}{2}$) for further studies. Also the stability
of the solution given in Eq's (\ref{genmet}), (\ref{sol}) and
(\ref{sk}) is an important issue of further investigation.

\acknowledgments

This work was supported by the research fund of Hanyang
University(HY-2005-1). The author would like to thank Dr. Gungwon
Kang(KISTI, Korea) for pointing out a crucial calculation mistake
in the original manuscript.

\end{document}